\documentclass[10pt]{article}
\usepackage[preprint]{tmlr}
\usepackage{amsmath,amssymb,booktabs,graphicx,tabularx,array}
\usepackage{hyperref}
\usepackage{url}
\usepackage{placeins}
\usepackage{needspace}
\hypersetup{hidelinks,pdfauthor={Han Chen, Yingrui Li},pdftitle={Do Evidence-Reading Diagnostics Improve Interface Selection in Small LLM Recommenders?}}
\input{glyphtounicode}
\title{Do Evidence-Reading Diagnostics Improve\\Interface Selection in Small LLM Recommenders?\thanks{LLM-based tools were used as assistive tools during this work. A full Tool assistance statement appears at the end of the main text.}}
\author{\name Han Chen \AND \name Yingrui Li}
\begin{document}
\maketitle
\begin{abstract}
Behavioral tests measure how a language model reads evidence. We ask whether those measurements help choose a recommendation interface. We evaluate six small instruction-tuned checkpoints across four recommendation domains with chronological evaluation and 3,426 evaluation users. Each request ranks eight candidates. A baseline selector chooses among history-only prompting, prompting with collaborative evidence, and score fusion. It uses observable features and six stability prompts that vary wording and candidate order. An augmented selector adds features from six evidence-reading prompts that ask the model to compare support counts. An interface chosen once on development (validation) data for each domain and checkpoint scores 0.5524 NDCG@5, compared with 0.5447 for the baseline selector and 0.5428 for the augmented selector. Adding the diagnostic features changes NDCG@5 by $-0.0019$ (95\% interval $[-0.0046,\,0.0004]$). The interval includes zero, and its upper bound is below the analysis plan's $0.005$ improvement target. Matching the selectors' hyperparameters also leaves the interval upper bound below that target. Evidence from retrieved similar users improves prompting by 0.0999 NDCG@5 over a control using randomly selected users matched for activity. The evidence-reading tests also reveal answer-position and tie-response biases. These results concern the tested selectors and candidate sets. They illustrate why diagnostic measurements should be evaluated by whether they improve recommendation choices beyond existing features and a fixed interface.
\end{abstract}

\section{Introduction}
LLM recommenders can receive collaborative information as text or combine their scores with a conventional recommender~\citep{hou2024rankers,pouryousef2026}. Choosing how to supply that information is a separate decision problem. Evidence-reading tests offer a plausible input to this choice: a model that correctly compares how many similar users support each candidate might use textual evidence more effectively. However, identifying the largest count and predicting a user's later preference require different judgments.

This distinction matters whenever behavioral measurements guide the choice of a model, prompt, or processing pipeline. A test can expose a weakness without identifying which alternative will work better for a particular request. Its value for selection depends on whether it adds useful information beyond the selector's existing inputs. Recommendation lets us study this question while keeping the language model fixed and changing how collaborative information is used.

We compare two selectors that use the same recommendation interfaces. The baseline uses observable features and stability measurements from alternative recommendation prompts. The augmented selector also uses responses to evidence-reading tests. Both selectors choose only after all interfaces have been scored, so the comparison evaluates the quality of the choice rather than saved inference work.

\paragraph{Related work.}
CheckList established systematic behavioral testing beyond aggregate task accuracy~\citep{ribeiro2020}. In recommendation, \citet{pouryousef2026} study how LLMs interpret collaborative signals and how prompting can improve their use. ReliGRec chooses prompts using a weak user-risk estimate~\citep{yang2026}. RouteRec reports that request-level hard selection remains below a fixed BM25 agent in its hit-rate comparison, whereas learned item-level aggregation recovers useful signal~\citep{zhou2026}. More broadly, request-level LLM routing also asks which option should handle a particular input: RouteLLM learns to route between stronger and weaker LLMs from preference data~\citep{ong2025routellm}, and RouterBench evaluates multi-LLM routing systems across models and tasks~\citep{hu2024routerbench}. Our setting keeps the checkpoint fixed and instead chooses among recommendation interfaces; the shared question is whether additional measurements improve per-request decisions beyond simpler baselines. BLaIR likewise distinguishes general embedding-benchmark performance from recommendation-specific utility~\citep{hou2026blair}.

Our results separate the usefulness of evidence from the usefulness of measuring how a model reads it. Retrieved evidence improves prompting relative to a random-user control, but adding reading features falls short of the targeted selection improvement. An interface chosen once on development data also outperforms both learned selectors. We examine the result through class-balanced diagnostic accuracy, feature variation across requests, matched selector hyperparameters, and user-level uncertainty. The practical comparison is therefore between a diagnostic-guided choice, a choice based on existing features, and a fixed interface. It asks what the diagnostic adds to the decision.

\section{Evaluation Design}
\subsection{Data and the controlled ranking task}
We use MovieLens 20M~\citep{harper2015}, Amazon Reviews 2023 Musical Instruments~\citep{hou2026blair}, KuaiRec~\citep{gao2022}, and Last.fm 1K~\citep{celma2010,lastfmweb}. We fit selectors and choose model settings on development data, then measure performance on a separate set of evaluation users. The development sets contain 498, 500, 500, and 350 users, respectively; the evaluation sets contain 1,000, 1,000, 1,000, and 426. No user belongs to both splits within a domain. MovieLens and Amazon reuse previously constructed cohorts with fixed candidate sets; Section~\ref{sec:inference} discusses their analysis history.

For KuaiRec and Last.fm, we sort events by timestamp and place global cutoffs at 80\% and 90\% of the event sequence. Histories contain distinct items with positive interactions before the relevant cutoff. We select the positive from the first subsequent timestamp group containing an unseen positive item. Prompts retain the latest eight distinct history items. KuaiRec uses videos from \texttt{big\_matrix.csv}, with a finite watch ratio of at least 1.0 defining a positive event. Last.fm uses artists and parseable listening events. Appendix~\ref{app:data} gives the exact cutoffs, eligibility rules, and preprocessing details.

Each request contains one later positive item and seven unseen alternatives matched for past popularity. We present the candidates in four fixed cyclic orders and score the next-token logits for the answer letters A--H. We map the resulting probabilities back to item identities and average across orders. A fixed SHA-256 item-key order breaks exact score ties. Input-order tie-breaking can otherwise make the metric depend on candidate storage order~\citep{guo2026ties}. If the positive has rank $r$, binary-relevance NDCG@5 is $1/\log_2(r+1)$ for $r\leq5$ and zero otherwise. Hit@1 is one exactly when $r=1$. These metrics describe \emph{eight-candidate ranking}. Their interpretation differs from full-catalog retrieval or online evaluation~\citep{krichene2020}.

We evaluate six pinned checkpoints: Qwen2.5-1.5B-Instruct, Qwen2.5-3B-Instruct, Qwen3-4B-Instruct-2507, Phi-4-mini-instruct, Gemma-3-4B-it, and Llama-3.2-3B-Instruct. A checkpoint is a fixed version of a model; Appendix~\ref{app:checkpoints} lists the repositories, and the supplement gives exact revisions. Qwen, Phi, Gemma, and Llama form the four analysis families. Users are the units for statistical inference. Checkpoints, rotations, candidate draws, and training seeds provide repeated measurements of those users.

\subsection{Interfaces and reference policies}
We call the three options \emph{interfaces}: history-only prompting (H), prompting with support counts from retrieved similar users (T), and score fusion (F). Fusion combines history-only and conventional recommender scores:
\begin{equation}
F=(1-\alpha)s(H)+\alpha s(C),
\end{equation}
Here $C$ is a user-based nearest-neighbor (user-KNN) score computed from past interactions, and $s$ applies midrank or z-score scaling. On development data, we choose the neighborhood size $K\in\{8,16,32,64\}$ for each domain. We choose the fusion weight $\alpha\in\{0,0.1,\ldots,1\}$ and scaling method for each domain/checkpoint pair. These settings remain fixed during evaluation.

On KuaiRec, five of six configurations select $\alpha=1$, so F uses only the user-KNN score. On Amazon, 840 of 1,000 evaluation requests have no retrieved support for any candidate. We retain these requests and also report results separately by support level. These conditions limit how often the interfaces can draw on complementary information.

Always-H, always-T, and always-F provide fixed references. We also report a post hoc reference called \emph{Dev-fixed}: for each domain/checkpoint pair, we choose the interface with the highest mean NDCG@5 on development data and use it for every evaluation request. Popularity, retrieved support, user-KNN, and five-seed BPR~\citep{rendle2009} use the same candidates. The three-interface oracle chooses the best observed H/T/F outcome for each request and checkpoint.

To test whether the source of evidence matters, we replace retrieved neighbors with distinct users drawn at random and matched for activity, while keeping the text prompt the same. This prompt control differs from the reference that ranks candidates directly by random-user support counts.

\subsection{Evidence-reading tests and selectors}
Separate development and evaluation test sets each contain 512 synthetic cases. They cover candidates with zero support, sparse support, equal maximum counts, one-count gaps, and near-ties when many candidates have support. Counts appear as digits, number words, or rows of user--item interactions. The group sizes of 8, 16, 32, and 64 refer to the number of synthetic neighbors. In the 254 direct cases, any item with the largest count is correct. The 258 pairwise cases compare two candidates and require an explicit tie answer when their counts are equal. Additional format variants measure response consistency. No case has zero support for every candidate.

Each recommendation request receives six evidence-reading prompts. Three ask directly which candidate has the most support, using digits, words, or interaction rows to show the same counts. Two compare a pair of candidates in opposite orders. The sixth is a \emph{zero/one check}: we select a maximum-support and a minimum-support candidate using deterministic tie-breaking, display their counts as 1 and 0, and ask which has greater support, with a tie option. The recommendation labels and candidate set remain fixed.

From the responses, we obtain 18 features measuring correctness against the displayed counts, probability assigned to valid answers, entropy, and agreement. Two more features give a test-coverage flag (zero for all-zero support, one otherwise) and reliability for the request's difficulty category, estimated on development tests. Together these form the 20-feature reading block. Given the fixed reliability lookup, the last two features depend on support counts and checkpoint identity already available to the baseline selector. We retain them in the fitted selectors but exclude them when summarizing variation in response-derived features.

The baseline selector, B, uses observable features $V$: support summaries, H/T scores, and a checkpoint indicator. It also uses stability features $S_6$ from three additional prompts for each of H and T. These \emph{stability prompts} test how scores change with wording and candidate order. Each pairs a different opening instruction with a left cyclic rotation of the same candidate permutation: 1, 3, or 5 positions, compared with 0, 2, 4, and 6 for main scoring. Histories, candidate metadata, and retrieved support counts stay fixed. For each interface, $S_6$ summarizes score entropy, the top-two margin, ranking agreement with the main scores, differences between the three score distributions, and top-choice agreement. Appendices~\ref{app:promptvariants} and~\ref{app:zeroone} give the exact prompts.

The augmented selector, B+R, adds the reading-feature block from the six evidence-reading prompts, denoted by $R_6$:
\begin{equation}
B: V+S_6 \qquad\text{versus}\qquad B+R:V+S_6+R_6.
\end{equation}
The baseline thus uses six additional diagnostic prompts, and the augmented selector uses twelve. Including the H/T scores needed by both selectors gives 14 versus 20 prompts per request/checkpoint, before scoring the random-evidence control. Both selectors run \emph{after all interfaces are scored}, so the comparison measures the quality of their choices rather than computation saved.

The selector uses a regression model, or \emph{gain forecaster}, to predict how much T and F improve utility $U$ over H. Its two targets are
\begin{equation}
g=\bigl(U(T)-U(H),\ U(F)-U(H)\bigr).
\end{equation}
The selector chooses the interface with the largest predicted gain, with H assigned a gain of zero. Exact ties use a deterministic ordering derived from SHA-256 keys for the domain, request, checkpoint, and interface.

We choose forecaster settings by five-fold cross-validation on development data, keeping all rows from a user in the same fold. We compare ridge regression and histogram gradient boosting by mean squared error (MSE) across the two targets. Each feature set selects its own configuration, which is then fitted on all development users. The selected cross-validated MSE is 0.128892 for B and 0.128903 for B+R. Their selected maximum-leaf counts are four and eight; both use minimum leaf size 80 and L2 penalty 10. We also fit the augmented selector using the baseline's hyperparameters. This \emph{matched-configuration sensitivity} checks whether the result changes when both feature sets use the same settings.

\subsection{Endpoints, inference, and analysis history}\label{sec:inference}
The primary endpoint is the change in NDCG@5 when the augmented selector replaces the baseline. We average checkpoint results within each user, users within each domain, and then the four domains with equal weight:
\begin{equation}
\Delta=\frac14\sum_{d=1}^4\frac1{n_d}\sum_{u=1}^{n_d}\frac16\sum_{m=1}^6
\left[U_{dum}(B+R)-U_{dum}(B)\right].
\end{equation}
Here $d$ indexes domains, $u$ users, $m$ checkpoints, and $n_d$ is the number of evaluation users in domain $d$. Positive values favor adding reading features. The analysis plan states this endpoint as a reduction in oracle regret; the expressions are identical because the same oracle score cancels in the difference. The practical improvement target is $0.005$. We measure reduction in the forecaster's two-target MSE separately.

For each domain, we resample users 10,000 times, keeping each user's two selector outcomes together. The primary pooled interval also resamples the four domains; this is the hierarchical bootstrap. We report a fixed-domain bootstrap alongside it, resampling users while holding the four domains fixed. One-sided user-level sign-flip tests assess improvement, with Holm correction across the four domain utility and four domain MSE comparisons. The intervals treat the candidates, checkpoints, and fitted selectors as fixed, so they exclude uncertainty from fitting. Each secondary table identifies its resampling scheme. Secondary intervals are nominal and do not provide simultaneous coverage.

We reuse MovieLens/Amazon evaluation cohorts from an earlier two-domain study for which aggregate results exist. We cannot establish whether those summaries were consulted before the extension plan was finalized. KuaiRec uses watch ratio $\geq1.0$, whereas an earlier feasibility record specifies $>2.0$; the reason and timing of that change are undocumented. We therefore interpret the September 18, 2026 plan as a specification of the reported analysis, without a verified separation between design decisions and earlier results. The intervals condition on the recorded procedures and exclude possible outcome-informed design uncertainty. Appendix~\ref{app:history} summarizes the supporting records.

The strong-positive rule requires at least $0.005$ pooled benefit, a 95\% interval excluding zero, positive effects in at least three domains, and no leave-domain/family reversal. The broader negative rule requires 90\% domain intervals within $[-0.005,0.005]$ in at least three domains, a non-ceiling reading test, and no single-family explanation. Both recorded rules are evaluated unchanged. Appendix~\ref{app:coverage} distinguishes original analyses from post hoc additions.

\clearpage
\section{Results}
\subsection{Evidence reading and recommendation choices}
\begin{figure}[!ht]
\centering
\includegraphics[width=0.76\textwidth]{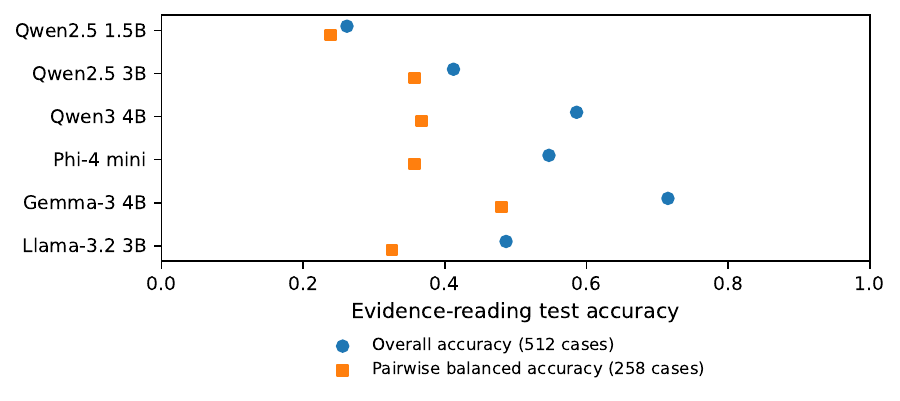}
\caption{Evidence-reading performance by checkpoint. Overall accuracy combines 254 direct and 258 pairwise evaluation cases. Pairwise balanced accuracy averages recall across left, right, and tie labels; its chance level is $1/3$. The two metrics have different denominators and scoring rules.}
\label{fig:audit}
\end{figure}

Overall evidence-reading accuracy ranges from 26.2\% to 71.5\% across checkpoints, combining direct and pairwise questions (Figure~\ref{fig:audit}). Pairwise class-balanced accuracy, which gives equal weight to left, right, and tie labels, ranges from 23.9\% to 48.0\%. Qwen3 chooses the first alternative on 95.0\% of pairwise cases; Gemma does so on 79.1\% and answers none of 54 ties correctly. Phi and Llama predict ties on 80.4\% and 74.5\% of non-ties. Aggregate accuracy therefore reflects count-comparison performance together with answer-position and tie-response tendencies.

Selection happens per request within a fixed checkpoint, so we examine how the reading features vary at that level. For most of the 18 response-derived features, variation is greater within checkpoints than between them. The median share of weighted variance explained by checkpoint identity is 10.0\%; three features exceed 50\%. Each feature varies within at least 12 of the 24 domain/checkpoint cells. This summary excludes the two input-derived covariates. The observed variation could still overlap with information in $V+S_6$ (Appendix~\ref{app:featurevariation}).

\begin{figure}[t]
\centering
\includegraphics[width=0.95\textwidth]{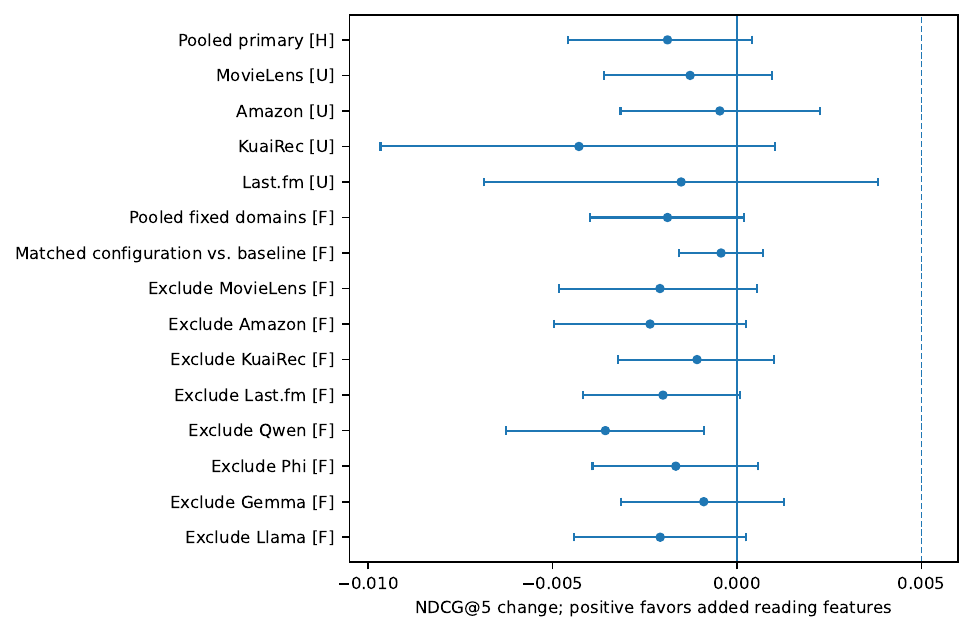}
\caption{NDCG@5 changes and 95\% intervals. H denotes domain-resampling hierarchical intervals; F denotes fixed-domain intervals; U denotes single-domain user-bootstrap intervals. Group exclusions use fixed-domain intervals, recomputing averages from the fitted policies' outcomes without refitting. The matched-configuration comparison is post hoc. The dashed line marks the $+0.005$ target.}
\label{fig:primary}
\end{figure}

The added reading features fall short of the improvement target: NDCG@5 changes by \textbf{$-0.0019$}, with a primary 95\% interval of \textbf{$[-0.0046,0.0004]$} and one-sided $p=0.9623$. All four domain point estimates are negative. The interval includes zero, and its upper bound is below the $0.005$ target. The post hoc fixed-domain 95\% interval is $[-0.0040,0.0002]$ (Figure~\ref{fig:primary}). Only MovieLens and Amazon have 90\% intervals entirely inside the broader rule's small-effect range, fewer than the required three domains. Both decision rules fail.

\begin{table}[!htbp]
\centering\small
\setlength{\tabcolsep}{3.0pt}
\caption{NDCG@5 on the same candidate sets. Dev-fixed chooses one H/T/F interface per domain/checkpoint using development data; this reference was added post hoc. B and B+R are the baseline and augmented selectors. BPR averages five seeds. The last row weights domains equally after averaging checkpoints within users.}
\label{tab:results}
\begin{tabular}{lrrrrrrrr}\toprule
Domain & H & T & F & Dev-fixed & B & B+R & Oracle & BPR\\\midrule
MovieLens & .514 & .636 & .669 & .669 & .670 & .669 & .769 & .698\\
Amazon & .509 & .513 & .523 & .536 & .532 & .532 & .619 & .452\\
KuaiRec & .352 & .424 & .458 & .458 & .441 & .436 & .641 & .546\\
Last.fm & .451 & .518 & .547 & .547 & .536 & .535 & .682 & .538\\
\midrule
Equal-domain & .456 & .523 & .549 & .552 & .545 & .543 & .678 & .559\\
\bottomrule\end{tabular}
\end{table}

\FloatBarrier
\subsection{Fixed-policy references and forecast error}
A fixed interface chosen on development data outperforms both selectors. Dev-fixed achieves \textbf{0.5524} NDCG@5, compared with \textbf{0.5447} for B and \textbf{0.5428} for B+R. Relative to Dev-fixed, the differences are $-0.0077$ ($[-0.0111,-0.0043]$) and $-0.0096$ ($[-0.0132,-0.0059]$), with fixed-domain intervals. Dev-fixed chooses fusion for 21 domain/checkpoint pairs and text for three Amazon pairs. B exceeds it only on MovieLens. BPR has the highest equal-domain mean among the non-oracle methods in Table~\ref{tab:results}, at 0.559, and exceeds every H/T/F interface on MovieLens and KuaiRec.

With the baseline's hyperparameters---a maximum of four leaves, minimum leaf size of 80, and L2 penalty of 10---the augmented selector scores 0.5443 NDCG@5: $-0.0004$ ($[-0.0016,0.0007]$) relative to B and $+0.0015$ ($[-0.0006,0.0035]$) relative to B+R. This fit uses no new hyperparameter search, and the upper bound for its improvement over B remains below the target.

Forecast error is slightly lower for the augmented selector: evaluation MSE falls from 0.1277 for B to 0.1274 for B+R, a relative reduction of 0.27\%. The absolute reduction is 0.0003 ($[-0.0004,0.0010]$). KuaiRec has the largest MSE improvement, 0.0011, but also the most negative utility change, $-0.0043$; its MSE result has Holm-adjusted $p=0.0648$.

\Needspace{7\baselineskip}
\subsection{Evidence controls and decision changes}
Evidence from retrieved similar users improves prompting relative to evidence from randomly selected users matched for activity. The two versions score 0.5229 and 0.4230 NDCG@5, respectively: a difference of \textbf{$+0.0999$} ($[0.0881,0.1114]$). The Hit@1 difference is $+0.1053$ ($[0.0913,0.1191]$), and all four domain point differences are positive. Both intervals hold domains fixed. Appendix~\ref{app:reviewchecks} reports the domain means.

The other feature comparisons also have negative NDCG@5 point estimates. Replacing six stability prompts with six reading prompts changes the score by $-0.0028$ ($[-0.0061,0.0003]$). Adding reading features to observable features alone gives $-0.0005$ ($[-0.0048,0.0031]$). Both intervals use hierarchical resampling. Hit@1 moves in the opposite direction for B+R versus B: $+0.0010$ ($[-0.0016,0.0037]$). Appendix~\ref{app:secondaries} reports these secondary comparisons. Relative to H, the baseline closes 39.9\% of the gap to the oracle, and the augmented selector closes 39.1\%; the appendix defines this \emph{oracle gap closed} measure.

The added features change which interface is selected in 22.93\% of weighted decisions. Across all decisions, 4.30\% improve NDCG@5, 4.76\% reduce it, and 13.87\% change the interface without changing utility. History-only choices rise from 6.91\% to 15.30\%, while fusion falls from 58.94\% to 49.87\%. The largest history-only increase is on Amazon, from 15.72\% to 33.65\%. The decision rule assigns H a predicted gain of zero, so it is selected when both competing gains are negative; ties at zero use the same hash-based rule. This describes how the fitted decisions change, but the reason for the prediction shifts remains unresolved.

\subsection{Support, domain, and timestamp sensitivities}
Removing requests with no retrieved support for any candidate leaves a negative NDCG@5 estimate: $-0.0025$ ($[-0.0053,0.0002]$). These requests include 218/1,000 MovieLens, 840/1,000 Amazon, 136/1,000 KuaiRec, and 33/426 Last.fm users. In KuaiRec and Last.fm, these requests account for about three-quarters of the negative estimated change in NDCG@5 (74.4\% and 74.6\%, respectively), although the synthetic test contains no all-zero cases. Those groups are small (136 and 33 users, respectively), and other requests also contribute.

The practical-target conclusion also holds in the post hoc domain subsets. With domains held fixed, the estimated change is $-0.0029$ ($[-0.0067,0.0008]$) for KuaiRec plus Last.fm, $-0.0011$ ($[-0.0032,0.0011]$) after excluding KuaiRec, and $-0.0009$ ($[-0.0027,0.0010]$) for the reused MovieLens/Amazon cohorts. Each upper bound is below $0.005$, as is the Last.fm-only upper bound of $0.0038$. These overlapping subsets use the same fitted selectors (Appendix~\ref{app:subsets}).

Excluding 29 Last.fm evaluation positives dated after the release's documented May 5, 2009 endpoint changes the pooled estimate by $+0.00015$ to $-0.0017$ (hierarchical 95\% interval $[-0.0045,0.0006]$), retaining an upper bound below $0.005$ (Appendix~\ref{app:timestamp_sensitivity}).

With two additional candidate draws for the same 250 users per domain, the changes are $-0.0045$ ($[-0.0103,0.0004]$) and $-0.0008$ ($[-0.0044,0.0028]$). Cross-fitting on the development cohort gives $-0.0008$ ($[-0.0041,0.0022]$). Among checkpoints, Gemma has the most negative effect, $-0.0068$; excluding it moves the pooled estimate toward zero, to $-0.0009$. Appendix~\ref{app:heterogeneity} reports the family summaries and the descriptive correlation between test accuracy and selection effects across the six checkpoints.

\section{Discussion and Limitations}
The study separates useful collaborative evidence from useful diagnostic features. Evidence from retrieved similar users improves prompting relative to the random-user control. Adding reading features, however, does not deliver the targeted $0.005$ improvement for the tested selectors. Both selectors also trail a fixed interface chosen on development data.

The results leave several explanations open. The tests ask models to compare visible support counts, whereas recommendation requires predicting later user preferences. Reading measurements may repeat information already in the score features, and answer-position tendencies can weaken their interpretation. The 18 response-derived features vary across requests within a checkpoint, but that variation may include useful signal, redundant information, or noise. The negative rule's non-ceiling condition concerns variation in test performance; it does not show that the measurements help choose an interface for individual requests.

The gain forecaster may also contribute to the result. Reading features enter both the T--H and F--H regressions, although F does not use the textual prompt. Those features could still reflect request difficulty relevant to H or the collaborative scores. Supplying them only to the T--H regression would test a different model specification. Another option is to start from Dev-fixed and switch interfaces only when the predicted gain exceeds a margin chosen on development data. Neither alternative was evaluated in this study.

The study covers controlled eight-candidate ranking, small pinned checkpoints, and four domains. Interaction histories respect the chronological cutoffs, but pretrained models and item metadata may contain later information. The synthetic test omits all-zero cases.

\section{Reproducibility}
The supplement reproduces primary and secondary statistics from saved paired outcomes, averaged across checkpoints within each user and stripped of original identifiers. Python and NumPy scripts cover comparisons with fixed interfaces and the random-user control, the fit using matched hyperparameters, and the subset analyses. Aggregate feature moments and confusion counts reproduce the diagnostic summaries. The scripts replay saved outcomes; they do not regenerate LLM predictions from source data. Appendix~\ref{app:replay} describes the execution environments, portability of model refitting, source-data checks, and incomplete computation records.

\section{Conclusion}
Across four domains and six checkpoints, evidence from retrieved similar users improves prompting relative to the random-user control. Adding features from six evidence-reading prompts does not deliver the plan's $0.005$ NDCG@5 improvement in interface selection, and a fixed interface chosen on development data outperforms both tested selectors. To judge the value of evidence-reading tests, measure whether their features improve recommendation choices beyond existing inputs and fixed interfaces. Report answer preferences and variation across requests within each checkpoint alongside aggregate accuracy.

\section*{Ethical Considerations}
Recommendation histories can expose sensitive preferences. The statistical supplement excludes original identifiers, histories, user-specific prompt text, and candidate lists; removing identifiers is not a formal privacy guarantee. Dataset and model redistribution terms still apply to upstream artifacts. Controlled ranking metrics leave fairness, user satisfaction, and deployment safety untested. Supplying collaborative histories to a model raises privacy issues independently of ranking utility.

\Needspace{5\baselineskip}
\paragraph{Tool assistance.}
The authors originated the main research question and the core study design. LLM-based tools, including ChatGPT and Codex, assisted with study-design discussions, code development, analysis execution, artifact verification, and manuscript drafting and revision. Claude and Gemini supplied draft critiques. These automated checks are distinct from independent human replication. The authors are responsible for the scientific claims, references, interpretation, and release decisions.

\clearpage
\bibliographystyle{tmlr}
\bibliography{references}
\clearpage
\appendix
\section{Data construction and analysis history}\label{app:data}
The study contains 1,848 development and 3,426 evaluation users. The splits are user-disjoint within each domain. KuaiRec items are videos; Last.fm items are artists.

\begin{table}[htbp]\centering
\caption{Cohort sizes and retrieved-evidence coverage. Zero support means zero retrieved counts for all eight candidates.}\label{tab:cohorts}
\begin{tabular}{lrrr}\toprule
Domain & Development & Evaluation & Zero support\\\midrule
MovieLens & 498 & 1000 & 218\\
Amazon & 500 & 1000 & 840\\
KuaiRec & 500 & 1000 & 136\\
Last.fm & 350 & 426 & 33\\\bottomrule
\end{tabular}\end{table}

\subsection{Preprocessing and candidate construction}
The KuaiRec/Last.fm preprocessing pipeline sorts events by timestamp. Cutoffs are the timestamps at indices $\lfloor0.8N\rfloor$ and $\lfloor0.9N\rfloor$. Development outcomes lie in the first resulting future window and evaluation outcomes in the second; histories use earlier positives. A binary user--item matrix deduplicates interactions, and prompts retain the latest eight distinct items. The positive comes from the earliest future timestamp group containing a previously unseen positive with past popularity. Seven alternatives are unseen in both the user's history and the positive items of that future window. Matching minimizes absolute distance in $\log(1+\mathrm{past\ popularity})$, followed by deterministic hash ordering and shuffling. Future-positive exclusion belongs to offline candidate construction; those labels are not selector inputs.

Neighbors use cosine similarity on binary past-positive histories, with at least five distinct history items. Up to 100 ranked overlapping users supply the 64 evidence peers. Random peers are distinct and matched by $\lfloor\log_2\mathrm{degree}\rfloor$ bins, with nearest-log-degree fallback. Outcome labels are stored separately from prompt inputs.

KuaiRec uses \texttt{KuaiRec 2.0/data/big\_matrix.csv}: 12,530,806 events from 7,176 users and 10,728 videos. Of these, 4,238,228 satisfy the implemented finite watch-ratio threshold of at least 1.0. Its exact global boundaries are Unix timestamps 1598650255 and 1598870571, with end-exclusive boundary 1599694333. Last.fm contains 19,098,862 parsed events from 992 users and 176,715 artists. Artist IDs identify items where available, with normalized artist names as fallback. Its exact boundaries are 1224212065, 1233604170, and 1380479525.

MovieLens and Amazon retain their earlier controlled-candidate inputs. The construction record excludes 100 noncontrolled development rows per domain and verifies zero split-user overlap. The analysis uses the preserved inputs and labels; the new-domain preprocessing above does not reconstruct their earlier source pipeline.

\subsection{Last.fm eligibility}
Feasibility records contain 726/884 raw interval users for development/evaluation, and 678/832 with at least five past artists. An initial allocation of 500 development users left only 310 evaluation users and was rejected before scoring. The final disjoint cohorts contain 350 and 426 users. Of these, 324/350 and 393/426 have nonzero retrieved support; the later positive is supported by the top 64 peers for 285/350 and 326/426. The records do not retain an exact per-filter attrition chain.

\subsection{Last.fm release dates and exclusion sensitivity}\label{app:timestamp_sensitivity}
The Last.fm 1K release describes listening histories through May 5, 2009~\citep{celma2010lastfmdata}. The parser accepts parseable ISO dates without a collection-date filter. An audit of the released data file found 153,473 events from 204 users on later UTC calendar dates. No flagged event entered prompt histories or retrieved/random-neighbor evidence: both evidence cutoffs precede May 6, 2009. However, 29 of 426 evaluation positives, and none of 350 development positives, are later-dated. The latest evaluated positive is June 19, 2009. The latest source timestamp is September 29, 2013; it was not a scored positive.

The distributed data file contains 19,098,862 rows, 52,006 fewer than the stated 19,150,868, despite matching the release's file identifier. All rows parse successfully; the discrepancy is between the data file and its documentation. The primary analyses retain all 426 Last.fm users; the following sensitivity excludes the 29 later-dated positives using saved outcomes. The supplement supplies the source audit and exclusion mask.

We exclude only evaluation users whose scored positive falls on a UTC calendar date later than May 5, 2009. The mask contains 29 exclusions and 397 retained Last.fm users, leaving 3,397 users across four domains. Six checkpoints are averaged within each user. Policies, candidate sets, labels, and the other three domains are unchanged.

Table~\ref{tab:timestamp_groups} reports paired outcomes for the full, retained, and flagged Last.fm groups. The flagged group's larger negative mean contributes to the full-sample estimate, but its removal does not recover the targeted pooled benefit. The identity $426\Delta_{\mathrm{all}}=397\Delta_{\mathrm{retained}}+29\Delta_{\mathrm{flagged}}$ holds to floating-point precision. Because each domain has weight $1/4$, the pooled shift is one quarter of the Last.fm shift.

\begin{table}[htbp]\centering\small
\caption{Last.fm timestamp groups. B denotes the baseline selector and B+R the augmented selector; $\Delta$ is their paired difference.}\label{tab:timestamp_groups}
\begin{tabular}{llrrrr}\toprule
Metric & Group & Users & B & B+R & $\Delta$\\\midrule
NDCG@5 & Full sample & 426 & 0.536223 & 0.534711 & $-0.001512$\\
NDCG@5 & Retained & 397 & 0.528338 & 0.527413 & $-0.000925$\\
NDCG@5 & Flagged & 29 & 0.644166 & 0.634623 & $-0.009543$\\
Hit@1 & Full sample & 426 & 0.288732 & 0.291080 & $+0.002347$\\
Hit@1 & Retained & 397 & 0.280856 & 0.283375 & $+0.002519$\\
Hit@1 & Flagged & 29 & 0.396552 & 0.396552 & $+0.000000$\\
\bottomrule\end{tabular}\end{table}

Table~\ref{tab:timestamp_intervals} applies both pooled resampling schemes. H resamples domains and users (10,000 draws); F fixes the four domains and resamples users (20,000 draws). U resamples Last.fm users only (10,000 draws). The full and restricted samples use matching random-number sequences. The primary and its decision rules remain unchanged.

\begin{table}[htbp]\centering\small
\caption{NDCG@5 timestamp sensitivity. H, F, and U are defined above. The retained-user comparisons are post hoc; full-sample rows provide the reference.}\label{tab:timestamp_intervals}
\begin{tabular}{llrrr}\toprule
Scope & Users retained & Scheme & $\Delta$ & 95\% interval\\\midrule
Four domains & 3426 & H & $-0.001881$ & $[-0.004580,0.000421]$\\
Four domains & 3397 & H & $-0.001735$ & $[-0.004530,0.000636]$\\
Four domains & 3426 & F & $-0.001881$ & $[-0.003971,0.000189]$\\
Four domains & 3397 & F & $-0.001735$ & $[-0.003896,0.000382]$\\
Last.fm & 426 & U & $-0.001512$ & $[-0.006856,0.003813]$\\
Last.fm & 397 & U & $-0.000925$ & $[-0.006422,0.004534]$\\
\bottomrule\end{tabular}\end{table}

The retained-user Hit@1 difference is $+0.001005$ pooled, with H interval $[-0.001655,0.003726]$ and F interval $[-0.001500,0.003511]$; Last.fm alone gives $+0.002519$ with U interval $[-0.003778,0.009236]$. The supplement gives the full and restricted results for both metrics.

\FloatBarrier
\Needspace{13\baselineskip}
\subsection{Recorded history and unresolved chronology}\label{app:history}
Table~\ref{tab:provenance} summarizes the study history and the remaining limits on confirmatory interpretation.

\begin{table}[htbp]\centering
\caption{Recorded analysis history and limits on confirmatory interpretation.}\label{tab:provenance}
\begin{tabularx}{\textwidth}{@{}p{.20\textwidth}X X@{}}\toprule
Issue & Documented facts & Remaining uncertainty\\\midrule
Reused outcomes & All 1,000 MovieLens and 1,000 Amazon evaluation inputs match earlier objects. The analysis uses the same labels; earlier outcome summaries exist. & Whether and when earlier summaries were consulted during extension design. The September 18, 2026 plan has no precise finalization timestamp.\\
KuaiRec positives & An earlier feasibility record specifies watch ratio $>2.0$. Both the preprocessing code and the scored cohorts use finite watch ratio $\geq1.0$. & The reason and timing of the change, and its relation to outcome inspection. No contemporaneous amendment was located.\\\bottomrule
\end{tabularx}\end{table}

The reported intervals account for sampling uncertainty under the recorded analysis. They exclude possible outcome-informed design uncertainty.

\section{Checkpoint and diagnostic specification}\label{app:checkpoints}
Table~\ref{tab:pins} lists the checkpoint repositories. The Qwen2.5 and Qwen3 checkpoints belong to one analysis family. Gemma and Llama use pinned mirrors whose weight tensors were checked against official source revisions. The supplement lists the exact revisions, which also determine tokenizers and configuration files.

\begin{table}[htbp]
\centering
\caption{Checkpoint repositories. Exact revisions are listed in the supplement.}\label{tab:pins}
\begin{tabularx}{\textwidth}{@{}p{0.23\textwidth}X@{}}\toprule
Checkpoint & Repository\\\midrule
Qwen2.5 1.5B & \nolinkurl{Qwen/Qwen2.5-1.5B-Instruct}\\
Qwen2.5 3B & \nolinkurl{Qwen/Qwen2.5-3B-Instruct}\\
Qwen3 4B & \nolinkurl{Qwen/Qwen3-4B-Instruct-2507}\\
Phi-4 mini & \nolinkurl{microsoft/Phi-4-mini-instruct}\\
Gemma-3 4B & \nolinkurl{unsloth/gemma-3-4b-it}\\
Llama-3.2 3B & \nolinkurl{unsloth/Llama-3.2-3B-Instruct}\\
\bottomrule\end{tabularx}\end{table}

\subsection{Test cases and observable request mapping}
Development and evaluation each contain 512 test cases. Equivalent representations are repeated measurements of the same case. The request-level reliability feature comes only from the development test set. A direct answer is valid when it names any maximum-support candidate; a pairwise answer must report a tie for equal counts. Accuracy scores the highest-probability answer; valid-answer mass instead sums probabilities over all admissible answers.

The outcome-independent mapping assigns the eight request support counts to a difficulty category by this precedence: tied maxima; at most one nonzero count; a unique maximum one count above the runner-up; a remaining case containing a zero; otherwise an all-positive case. All other all-positive requests enter the final category, including requests without near-tied maxima. All-zero cases map to tied maxima and have test-coverage flag zero; the flag is one for every nonzero support vector. The test set contains positive tied maxima but no all-zero case. The mapping uses visible support counts.

\begin{table}[htbp]
\centering
\caption{Held-out test accuracy. Overall n=512, direct n=254, and pairwise n=258. The two decision formats use different primary cases.}\label{tab:auditnumbers}
\begin{tabular}{lrrr}\toprule
Checkpoint & Overall & Direct & Pairwise\\\midrule
Qwen2.5 1.5B & 26.17\% & 30.71\% & 21.71\%\\
Qwen2.5 3B & 41.21\% & 41.34\% & 41.09\%\\
Qwen3 4B & 58.59\% & 72.83\% & 44.57\%\\
Phi-4 mini & 54.69\% & 82.28\% & 27.52\%\\
Gemma-3 4B & 71.48\% & 85.43\% & 57.75\%\\
Llama-3.2 3B & 48.63\% & 72.83\% & 24.81\%\\
\bottomrule\end{tabular}\end{table}

\FloatBarrier
\subsection{Recommendation stability prompts}\label{app:promptvariants}
The same prompt construction is used in all four domains. Each request has one deterministic base permutation of its eight candidates, determined by the domain, request identifier, and a fixed prompt seed. A rotation of $r$ positions moves the first $r$ candidates to the end. Main H, T, and random-evidence prompts use rotations 0, 2, 4, and 6 with the opening instruction:
\begin{quote}
Select the candidate this user is most likely to enjoy next.
\end{quote}
The three additional prompts for each of H and T use the following fixed instruction--rotation pairs:
\begin{description}
\item[Rotation 1:] Which candidate best matches what this user is likely to prefer next?
\item[Rotation 3:] Choose the item most likely to receive a positive response from this user.
\item[Rotation 5:] Recommend the candidate that is the strongest next-item fit for this user.
\end{description}
Wording and order change together; they are not fully crossed. All variants preserve the same history, eight candidates, and item metadata. T also preserves each item's retrieved support count, remapped to its displayed letter, and the evidence description: ``The community evidence comes from 64 past-only similar users.'' Each prompt ends with ``Reply with only the selected letter.'' These six additional prompts supply $S_6$; the random-evidence control has only its four main prompts.

\subsection{Zero/one check}\label{app:zeroone}
The zero/one check uses the request's rotation-0 candidate labels. It sorts those labels by descending observed support, breaking ties with a deterministic key derived from the request identifier and displayed letter. The supplement gives the exact key construction.
Let $a$ be the first label and $b$ the last. They identify a maximum-support and a minimum-support candidate; among ties, $a$ has the smallest key and $b$ the largest. The labels are distinct even when all eight counts are equal. The prompt replaces their displayed counts with 1 and 0 and omits the other six candidates:
\begin{quote}\small\ttfamily\raggedright
This is a local diagnostic counterfactual and is not recommendation evidence.\\
Candidate \{a\} has one supporting neighbor.\\
Candidate \{b\} has zero supporting neighbors.\\
Return exactly \{a\} if it has greater support, \{b\} if it has greater support, or T if tied.
\end{quote}
The only valid answer is $a$, since $1>0$, regardless of the observed counts. The scorer compares next-token probabilities over $a$, $b$, and the tie token \texttt{T}, recording the probability assigned to $a$ and whether the highest-probability answer is $a$. These two measurements enter the reading-feature block. The prompt contains no user history or item metadata and leaves the recommendation inputs and outcomes unchanged. Exactly one such prompt is scored per request.

\section{Development-selected settings}\label{app:settings}
Neighborhood sizes are 64, 64, 8, and 64 for MovieLens, Amazon, KuaiRec, and Last.fm, respectively. The fusion coefficient multiplies the collaborative score, so the boundary $\alpha=1$ is user-KNN alone. Table~\ref{tab:fusion} retains all 24 checkpoint--domain choices. M denotes midrank scaling and Z denotes z-score scaling.

\begin{table}[htbp]\centering
\caption{Fusion coefficient and scaling selected from development utilities and applied unchanged to evaluation.}\label{tab:fusion}
\begin{tabular}{lrrrr}\toprule
Checkpoint & MovieLens & Amazon & KuaiRec & Last.fm\\\midrule
Qwen2.5 1.5B & 0.8 Z & 0.7 Z & 1.0 M & 0.7 Z\\
Qwen2.5 3B & 0.7 Z & 0.4 M & 0.5 M & 0.9 Z\\
Qwen3 4B & 0.8 Z & 0.6 M & 1.0 M & 0.7 Z\\
Phi-4 mini & 0.9 Z & 0.6 Z & 1.0 M & 0.8 Z\\
Gemma-3 4B & 0.8 Z & 0.6 Z & 1.0 M & 0.7 Z\\
Llama-3.2 3B & 0.8 Z & 0.6 Z & 1.0 M & 0.7 Z\\
\bottomrule\end{tabular}\end{table}

The gain-forecast search considers ridge penalties 0.01, 0.1, 1, 10, and 100, and separate histogram-gradient-boosting regressors for the two gain targets. Boosting candidates use maximum leaves 4, 8, or 16; minimum leaf sizes 20, 40, or 80; L2 penalties 1 or 10; learning rate 0.05; and 120 iterations. Five-fold development cross-validation groups all checkpoint rows from each user. The baseline and augmented selected fits have maximum leaves 4 and 8, respectively, and both use minimum leaf size 80 and L2 penalty 10. Hyperparameters are selected separately for the two feature sets. BPR results average five seeded fits.

\begin{table}[htbp]\centering
\caption{Development-selected cross-validation MSE and configurations. The mapping to archived feature keys is in the supplement README.}\label{tab:devselection}
\begin{tabular}{lrrrr}\toprule
Inputs & Max. leaves & Min. leaf & L2 & CV MSE\\\midrule
Observable only & 4 & 80 & 10 & 0.129140\\
Observable + reading & 8 & 80 & 1 & 0.129167\\
B & 4 & 80 & 10 & 0.128892\\
B+R & 8 & 80 & 10 & 0.128903\\
\bottomrule\end{tabular}\end{table}

The best B+R configuration improves over its own four-leaf, min-leaf-80, L2-10 candidate by approximately $2.6\times10^{-5}$ development MSE. The post hoc matched-configuration fit uses B's selected settings with augmented inputs. It fits exactly two target regressors on all 11,088 development checkpoint rows, in the same row order, with learning rate 0.05, 120 iterations, early stopping disabled, and fixed target-specific random states. The fit uses development data only, with no additional search. Results are in Appendix~\ref{app:reviewchecks}; the selected policies remain unchanged.

\section{Numerical inference and decision rules}\label{app:inference}
\paragraph{Scope of secondary analyses.} All secondary intervals are nominal and condition on the fitted policies. Overlapping subgroups share users and checkpoints. Each table or accompanying paragraph identifies its resampling scheme.

The primary and four domain utility results are listed in Table~\ref{tab:op}. The hierarchical pooled bootstrap and single-domain intervals use 10,000 draws. Seeds, random-number ordering, and file hashes are listed in the supplement. The one-sided tests target positive improvement. The eight domain tests comprise four utility and four two-gain-MSE comparisons; Holm-adjusted $p$-values govern the domain-specific improvement tests.

\begin{table}[htbp]\centering
\caption{Change in selected NDCG@5 and interval containment. All four domain utility Holm-adjusted $p$-values equal 1.000. The pooled one-sided $p$-value is 0.962304.}\label{tab:op}
\begin{tabular}{lrrr}\toprule
Domain & Mean & 95\% interval & 90\% interval\\\midrule
MovieLens 20M & $-0.001269$ & $[-0.003592,0.000960]$ & $[-0.003222,0.000622]$\\
Amazon Instruments & $-0.000463$ & $[-0.003154,0.002247]$ & $[-0.002742,0.001832]$\\
KuaiRec & $-0.004282$ & $[-0.009660,0.001038]$ & $[-0.008851,0.000237]$\\
Last.fm 1K & $-0.001512$ & $[-0.006856,0.003813]$ & $[-0.006067,0.002928]$\\
Pooled & $-0.001881$ & $[-0.004580,0.000421]$ & ---\\
\bottomrule\end{tabular}\end{table}

Only MovieLens and Amazon meet the 90\% interval-containment condition within $[-0.005,0.005]$. The required count is three domains. This failure is sufficient to reject the protocol's broader-negative designation, irrespective of its other conditions. The positive designation also fails. 

\begin{table}[htbp]\centering
\caption{Reduction in two-target gain-forecast MSE. The last column reports the eight-test Holm correction.}\label{tab:mse}
\begin{tabular}{lrrr}\toprule
Domain & Mean & 95\% interval & Holm $p$\\\midrule
MovieLens 20M & $-0.000142$ & $[-0.001382,0.001067]$ & 1.000000\\
Amazon Instruments & $0.000364$ & $[-0.000054,0.000787]$ & 0.299570\\
KuaiRec & $0.001095$ & $[0.000201,0.001998]$ & 0.064794\\
Last.fm 1K & $0.000054$ & $[-0.001290,0.001451]$ & 1.000000\\
Pooled & $0.000343$ & $[-0.000362,0.001022]$ & ---\\
\bottomrule\end{tabular}\end{table}

Group exclusions recompute averages from the existing fitted outcomes. The two additional candidate draws reuse 250 evaluation users per domain. Cross-fitting uses the development cohort that selected settings.

\begin{table}[htbp]\centering
\caption{Pooled interval sensitivity. Hierarchical domain resampling is the primary; the fixed-domain interval is post hoc. Both operate on user-level checkpoint averages.}\label{tab:fixedbootstrap}
\begin{tabular}{lrrr}\toprule
Resampling & Mean & 95\% interval & 90\% interval\\\midrule
Primary hierarchical & $-0.001881$ & $[-0.004580,0.000421]$ & $[-0.004091,0.000085]$\\
Fixed domains; post hoc & $-0.001881$ & $[-0.003971,0.000189]$ & $[-0.003618,-0.000118]$\\
\bottomrule\end{tabular}\end{table}

The fixed-domain sensitivity uses 20,000 draws, resampling users within domains. The primary hierarchical method additionally resamples four domain labels. Both estimate the same equal-domain mean but represent different sources of uncertainty; extrapolation from four domains remains limited.

\subsection{Domain-subset sensitivities}\label{app:subsets}
For Last.fm alone, we use the 10,000-draw user-bootstrap interval from Table~\ref{tab:op}. The three multi-domain restrictions use 20,000 fixed-domain paired-user draws, with equal domain weights and shared within-domain draws across overlapping subsets.

\begin{table}[htbp]\centering
\caption{Domain restrictions. Last.fm uses the interval in Table~\ref{tab:op}; the other rows are post hoc fixed-domain user-bootstrap intervals.}\label{tab:subsets}
\begin{tabular}{lrrr}\toprule
Domains & Users & Mean & 95\% interval\\\midrule
Last.fm & 426 & $-0.001512$ & $[-0.006856,0.003813]$\\
KuaiRec + Last.fm & 1426 & $-0.002897$ & $[-0.006713,0.000842]$\\
MovieLens + Amazon + Last.fm & 2426 & $-0.001081$ & $[-0.003231,0.001055]$\\
MovieLens + Amazon & 2000 & $-0.000866$ & $[-0.002651,0.000957]$\\
\bottomrule\end{tabular}\end{table}
Every reported upper bound is below $0.005$. These restrictions describe changes in evaluation membership while retaining the same fitted selectors; they do not establish independent pre-outcome study design.

\section{Secondary endpoints and reference policies}\label{app:secondaries}
These secondary analyses use saved predictions and policies. The plan named the endpoints but left some reporting formulas and bootstrap seeds unspecified; those choices are recorded in the replay scripts. Table~\ref{tab:ablations} uses hierarchical domain-resampling intervals throughout.

\begin{table}[htbp]\centering
\caption{Conventional references, fixed interfaces, and primary selectors. Each cell is NDCG@5 / Hit@1 on identical candidates. Random support ranks candidates by random-neighbor counts; the LLM prompt control appears in Appendix~\ref{app:reviewchecks}.}\label{tab:allbaselines}
\begin{tabular}{lrrrr}\toprule
Method & MovieLens & Amazon & KuaiRec & Last.fm\\\midrule
Popularity & 0.3767 / 0.0750 & 0.3702 / 0.0990 & 0.3563 / 0.0490 & 0.3750 / 0.1197\\
Retrieved support & 0.6094 / 0.3870 & 0.4208 / 0.1860 & 0.4579 / 0.2020 & 0.5208 / 0.2864\\
Random support & 0.4025 / 0.1650 & 0.3841 / 0.1370 & 0.3945 / 0.1370 & 0.3234 / 0.1080\\
User-KNN & 0.6111 / 0.3880 & 0.4203 / 0.1860 & 0.4585 / 0.1890 & 0.5263 / 0.2981\\
BPR (five seeds) & 0.6978 / 0.4496 & 0.4521 / 0.1926 & 0.5461 / 0.2750 & 0.5384 / 0.2761\\
H & 0.5138 / 0.2720 & 0.5087 / 0.2468 & 0.3515 / 0.1195 & 0.4513 / 0.1960\\
T & 0.6359 / 0.4012 & 0.5130 / 0.2755 & 0.4243 / 0.1773 & 0.5185 / 0.2762\\
F & 0.6689 / 0.4343 & 0.5226 / 0.2648 & 0.4581 / 0.1905 & 0.5466 / 0.2942\\
B & 0.6701 / 0.4398 & 0.5320 / 0.2852 & 0.4406 / 0.1997 & 0.5362 / 0.2887\\
B+R & 0.6688 / 0.4403 & 0.5316 / 0.2863 & 0.4363 / 0.1995 & 0.5347 / 0.2911\\
\bottomrule\end{tabular}\end{table}

\begin{table}[htbp]\centering
\caption{Feature-set comparisons. Positive effects favor the first-listed policy. All rows are secondary except the unchanged primary listed for context. The equal-prompt-count comparison uses six diagnostic prompts on each side. All intervals here use the hierarchical domain-resampling scheme.}\label{tab:ablations}
\begin{tabular}{lrr}\toprule
Contrast & Effect & 95\% interval\\\midrule
B+R minus B (primary) & $-0.001881$ & $[-0.004580,0.000421]$\\
Reading + observable minus observable & $-0.000529$ & $[-0.004761,0.003052]$\\
B minus observable & $0.002241$ & $[-0.000042,0.004204]$\\
Reading + observable minus B (equal-prompt-count) & $-0.002770$ & $[-0.006147,0.000303]$\\
B+R minus reading + observable & $0.000889$ & $[-0.002034,0.004027]$\\
B+R minus B (Hit@1) & $0.000962$ & $[-0.001643,0.003701]$\\
\bottomrule\end{tabular}\end{table}

All comparisons in this table use 10,000 hierarchical draws. Hit@1 is recovered from binary one-positive NDCG utility: utility equals one if and only if the positive is ranked first. All observed saved arm utilities were checked against the six allowed values for ranks 1--5 and rank greater than 5.

\begin{table}[htbp]\centering
\caption{Fixed-fusion reference and oracle gap closed relative to H. Always-F differs from Dev-fixed.}\label{tab:fixedreference}
\begin{tabular}{lrrrr}\toprule
Domain & B minus F & B+R minus F & B gap closed & B+R gap closed\\\midrule
MovieLens & $0.001113$ & $-0.000155$ & 61.31\% & 60.81\%\\
Amazon & $0.009447$ & $0.008984$ & 21.22\% & 20.80\%\\
KuaiRec & $-0.017562$ & $-0.021844$ & 30.77\% & 29.29\%\\
Last.fm & $-0.010416$ & $-0.011927$ & 36.77\% & 36.12\%\\
Equal-domain & $-0.004354$ & $-0.006236$ & 39.94\% & 39.09\%\\
\bottomrule\end{tabular}\end{table}

Oracle gap closed measures how much of the difference between H and the oracle is covered by a policy $\pi$: $(\bar U(\pi)-\bar U(H))/(\bar U(\mathrm{oracle})-\bar U(H))$. To obtain the means, we average checkpoints within users and then users within each domain. The pooled row takes the ratio after averaging domains equally. The separate Dev-fixed reference chooses its interface using development utilities only; Appendix~\ref{app:reviewchecks} reports its evaluation results and uncertainty. 

\begin{table}[htbp]\centering
\caption{Chosen-interface frequencies (percent). Users are averaged within domains and domains equally for the last row; checkpoints are repeated decisions.}\label{tab:choices}
\begin{tabular}{lrrrrrr}\toprule
Domain & B: H & B: T & B: F & B+R: H & B+R: T & B+R: F\\\midrule
MovieLens & 3.87 & 26.57 & 69.57 & 10.38 & 28.48 & 61.13\\
Amazon & 15.72 & 53.53 & 30.75 & 33.65 & 48.70 & 17.65\\
KuaiRec & 5.07 & 25.55 & 69.38 & 11.38 & 29.98 & 58.63\\
Last.fm & 2.97 & 30.99 & 66.04 & 5.79 & 32.16 & 62.05\\
Equal-domain & 6.91 & 34.16 & 58.94 & 15.30 & 34.83 & 49.87\\
\bottomrule\end{tabular}\end{table}

\subsection{Calibration and evidence difficulty}
We assess evidence-reading answer probabilities using ten-bin expected calibration error (ECE) and a Brier score over the admissible answers. When several items share the maximum count in a direct question, the target probability is divided equally among them. We separately summarize calibration of the gain forecasts with a post hoc linear analysis that weights domains equally, users equally within domains, and checkpoints equally within users. The supplement gives the intercepts, slopes, and mean prediction biases.

\begin{table}[htbp]\centering
\caption{Held-out evidence-reading calibration, 512 primary cases per checkpoint. ECE uses ten confidence bins; different task formats are combined in the reported overall score.}\label{tab:auditcal}
\begin{tabular}{lrrrr}\toprule
Checkpoint & Accuracy & Mean confidence & ECE & Brier\\\midrule
Qwen2.5 1.5B & 0.2617 & 0.6178 & 0.3583 & 0.9730\\
Qwen2.5 3B & 0.4121 & 0.8190 & 0.4096 & 0.9647\\
Qwen3 4B & 0.5859 & 0.9652 & 0.3854 & 0.8224\\
Phi-4 mini & 0.5469 & 0.7647 & 0.2298 & 0.6789\\
Gemma-3 4B & 0.7148 & 0.9847 & 0.2730 & 0.5998\\
Llama-3.2 3B & 0.4863 & 0.7148 & 0.2456 & 0.7449\\
\bottomrule\end{tabular}\end{table}

\begin{table}[htbp]\centering
\caption{Test accuracy by synthetic difficulty category. Tie n=104; all other difficulty categories combined n=408; zero-mixed n=100, sparse n=103, one-count n=103, dense-near-tie n=102. The difficulty category combines 50 direct and 54 pairwise cases; only the latter are explicit pairwise tie answers.}\label{tab:regimeaudit}
\begin{tabular}{lrrrrrr}\toprule
Checkpoint & Tie & Other & Zero-mixed & Sparse & One-count & Dense\\\midrule
Qwen2.5 1.5B & 0.4135 & 0.2230 & 0.1700 & 0.2136 & 0.2524 & 0.2549\\
Qwen2.5 3B & 0.3462 & 0.4289 & 0.4300 & 0.5728 & 0.3204 & 0.3922\\
Qwen3 4B & 0.4135 & 0.6299 & 0.6600 & 0.7282 & 0.5728 & 0.5588\\
Phi-4 mini & 0.8750 & 0.4632 & 0.4600 & 0.5049 & 0.5049 & 0.3824\\
Gemma-3 4B & 0.4615 & 0.7794 & 0.8700 & 0.9223 & 0.6796 & 0.6471\\
Llama-3.2 3B & 0.7981 & 0.4069 & 0.3900 & 0.4951 & 0.3786 & 0.3627\\
\bottomrule\end{tabular}\end{table}

Aggregate difficulty category accuracy combines decision formats. Appendix~\ref{app:ties} supplies the actual label-by-response cross-tab; it shows both tie avoidance and excessive tie responses, depending on checkpoint.

\begin{table}[htbp]\centering
\caption{Real-request difficulty breakdown: each cell is user count / mean primary NDCG@5 effect. The fixed precedence partitions requests; all-zero cases are split from positive tied maxima for readability.}\label{tab:realregime}
\begin{tabular}{lrrrr}\toprule
Category & MovieLens & Amazon & KuaiRec & Last.fm\\\midrule
All zero & 218 / +0.0034 & 840 / +0.0005 & 136 / -0.0234 & 33 / -0.0146\\
Tied maximum (>0) & 97 / -0.0004 & 15 / -0.0113 & 128 / -0.0057 & 76 / +0.0076\\
Sparse & 126 / +0.0023 & 132 / -0.0047 & 111 / -0.0118 & 41 / -0.0016\\
One-count gap & 132 / -0.0001 & 8 / -0.0128 & 169 / -0.0034 & 80 / -0.0085\\
Other mixed-zero & 320 / -0.0052 & 5 / +0.0000 & 203 / -0.0025 & 61 / +0.0009\\
Other all-positive & 107 / -0.0054 & 0 / --- & 253 / +0.0080 & 135 / -0.0004\\
\bottomrule\end{tabular}\end{table}

The real-request breakdown reconstructs counts from the saved observable-feature vector, where eight integer support counts divided by 64 occupy zero-based indices 54--61. Integer recovery, all-zero flags, and six measurements per user are checked. The fixed mapping is invariant to the history-based ordering of those eight counts.

\section{Fixed-policy and specificity comparisons}\label{app:reviewchecks}
These analyses were added after the original analysis. The fixed-arm reference and matched forecaster are post hoc specifications. Random-evidence prompting was listed in the plan and already scored, but the reported paired-interval specification was added after the original analysis. Each new interval uses 10,000 within-domain resamples of users after checkpoint averaging, holds the four domains fixed, and weights their means equally. All intervals in this section use that fixed-domain scheme.

\subsection{Development-selected fixed interface}
The fixed reference maximizes mean development NDCG@5 independently in each of 24 domain/checkpoint cells. The analysis specifies H/T/F order for exact ties, but no such tie occurs. It selects T for Amazon with Qwen3 4B, Phi-4 mini and Gemma-3 4B, and F in all remaining cells. Full development utilities are supplied in \texttt{DEVELOPMENT\_FIXED\_ARM\_SELECTION.csv}.

\begin{table}[htbp]\centering\small
\caption{Evaluation performance of the development-selected fixed policy, compared with the two selectors. Each cell is NDCG@5 / Hit@1.}\label{tab:devfixed}
\begin{tabular}{lrrr}\toprule
Domain & Dev-fixed & B & B+R\\\midrule
MovieLens & 0.668944 / 0.434333 & 0.670058 / 0.439833 & 0.668789 / 0.440333\\
Amazon & 0.535911 / 0.285667 & 0.532019 / 0.285167 & 0.531556 / 0.286333\\
KuaiRec & 0.458119 / 0.190500 & 0.440557 / 0.199667 & 0.436275 / 0.199500\\
Last.fm & 0.546638 / 0.294210 & 0.536223 / 0.288732 & 0.534711 / 0.291080\\
\midrule Equal-domain & 0.552403 / 0.301177 & 0.544714 / 0.303350 & 0.542833 / 0.304312\\\bottomrule\end{tabular}\end{table}

\begin{table}[htbp]\centering
\caption{Paired fixed-policy comparisons, conditional on the four domains. The post hoc intervals use fixed-domain resampling.}\label{tab:devfixeddelta}
\begin{tabular}{lrr}\toprule
Contrast & Difference & 95\% interval\\\midrule
B minus fixed (NDCG@5) & $-0.007689$ & $[-0.011056,-0.004270]$\\
B+R minus fixed (NDCG@5) & $-0.009570$ & $[-0.013204,-0.005897]$\\
B minus fixed (Hit@1) & $0.002172$ & $[-0.002023,0.006299]$\\
B+R minus fixed (Hit@1) & $0.003134$ & $[-0.001389,0.007622]$\\
\bottomrule\end{tabular}\end{table}

Fixed-arm NDCG@5 advantages do not extend to a demonstrated Hit@1 advantage: both selector-minus-fixed Hit@1 point differences are positive and their intervals include zero.

\subsection{Activity-matched random-evidence prompts}
The control uses the same text-prompt interface with activity-matched random-neighbor evidence, averaged across four saved main rotations. Here random-neighbor counts enter the LLM prompt; the random-support baseline ranks directly by those counts. Table~\ref{tab:randomprompt} shows domain means; the pooled paired NDCG@5 difference is $+0.099896$ with interval $[0.088104,0.111432]$, and the Hit@1 difference is $+0.105346$ with interval $[0.091272,0.119068]$. Retrieved evidence outperforms this prompt control in every domain.

\begin{table}[htbp]\centering\small
\caption{Retrieved (T) and activity-matched random-neighbor evidence prompts. The last row averages the four domain means equally.}\label{tab:randomprompt}
\begin{tabular}{lrrrrrr}\toprule
Domain & T NDCG & Random NDCG & Difference & T Hit@1 & Random Hit@1 & Difference\\\midrule
MovieLens & 0.635868 & 0.481366 & 0.154502 & 0.401167 & 0.229000 & 0.172167\\
Amazon & 0.513007 & 0.486436 & 0.026572 & 0.275500 & 0.234500 & 0.041000\\
KuaiRec & 0.424308 & 0.363141 & 0.061168 & 0.177333 & 0.126000 & 0.051333\\
Last.fm & 0.518488 & 0.361146 & 0.157342 & 0.276213 & 0.119327 & 0.156886\\
\midrule Equal-domain & 0.522918 & 0.423022 & 0.099896 & 0.282553 & 0.177207 & 0.105346\\\bottomrule\end{tabular}\end{table}

\subsection{Matched-configuration sensitivity}
The matched augmented policy scores 0.544284 NDCG@5 and 0.303669 Hit@1, compared with 0.544714/0.303350 for B and 0.542833/0.304312 for B+R. Table~\ref{tab:matched} reports all contrasts from this fit. MSE-reduction rows subtract matched MSE from the comparator MSE, so positive values favor the matched fit. Matching removes the hyperparameter difference while allowing the fitted functions to change with the inputs.

\begin{table}[htbp]\centering
\caption{Post hoc matched-configuration results with fixed-domain intervals. No new configuration search was performed.}\label{tab:matched}
\begin{tabular}{lrr}\toprule
Contrast & Difference & 95\% interval\\\midrule
Matched minus B (NDCG@5) & $-0.000430$ & $[-0.001581,0.000706]$\\
Matched minus B+R (NDCG@5) & $0.001451$ & $[-0.000594,0.003495]$\\
Matched minus B (Hit@1) & $0.000319$ & $[-0.001018,0.001670]$\\
Matched minus B+R (Hit@1) & $-0.000643$ & $[-0.003108,0.001775]$\\
B minus matched (MSE) & $-0.00000827$ & $[-0.00008854,0.00007221]$\\
B+R minus matched (MSE) & $-0.00035110$ & $[-0.00088771,0.00017566]$\\
\bottomrule\end{tabular}\end{table}

Per-domain effects are supplied in the supplement.

\section{Response-format and feature-variation diagnostics}\label{app:ties}
Actual pairwise ties are identified by the saved valid-answer label being the explicit tie answer. Direct cases with several admissible maxima are identified by the number of valid answers. The primary test set contains 54 pairwise ties and 204 non-ties, plus 50 direct multiple-maxima and 204 unique-maximum cases. The 1,024 additional format variants per checkpoint are excluded. The label-derived pairwise-tie subset coincides with the synthetic tie difficulty category in this test set.

\begin{table}[htbp]\centering\small
\caption{Actual label-based accuracies and false tie responses (percent). Pairwise denominators are 54 ties and 204 non-ties; direct denominators are 50 multiple maxima and 204 unique maxima.}\label{tab:actualties}
\begin{tabular}{lrrrrr}\toprule
 & \multicolumn{3}{c}{Pairwise} & \multicolumn{2}{c}{Direct}\\
Checkpoint & Tie acc. & Non-tie acc. & False tie & Multiple & Unique\\\midrule
Qwen2.5 1.5B & 35.19 & 18.14 & 69.12 & 48.00 & 26.47\\
Qwen2.5 3B & 11.11 & 49.02 & 19.61 & 60.00 & 36.76\\
Qwen3 4B & 1.85 & 55.88 & 0.98 & 84.00 & 70.10\\
Phi-4 mini & 81.48 & 13.24 & 80.39 & 94.00 & 79.41\\
Gemma-3 4B & 0.00 & 73.04 & 1.47 & 96.00 & 82.84\\
Llama-3.2 3B & 74.07 & 11.76 & 74.51 & 86.00 & 69.61\\
\bottomrule\end{tabular}\end{table}

\begin{table}[htbp]\centering
\caption{Pairwise confusion counts. Each triple is predicted (left, right, tie); columns index the true answer. Each checkpoint has the same 106 left, 98 right, and 54 tie labels.}\label{tab:tieconfusion}
\begin{tabular}{lrrr}\toprule
Checkpoint & True left & True right & True tie\\\midrule
Qwen2.5 1.5B & (17, 15, 74) & (11, 20, 67) & (20, 15, 19)\\
Qwen2.5 3B & (77, 8, 21) & (56, 23, 19) & (38, 10, 6)\\
Qwen3 4B & (104, 0, 2) & (88, 10, 0) & (53, 0, 1)\\
Phi-4 mini & (24, 3, 79) & (10, 3, 85) & (7, 3, 44)\\
Gemma-3 4B & (104, 1, 1) & (51, 45, 2) & (49, 5, 0)\\
Llama-3.2 3B & (14, 8, 84) & (20, 10, 68) & (12, 2, 40)\\
\bottomrule\end{tabular}\end{table}

Gemma's direct multiple-maxima accuracy is 96\%, but its explicit pairwise-tie accuracy is zero. Phi and Llama exhibit the opposite problem of frequent tie responses on non-ties. The checkpoints therefore have distinct answer tendencies. 

\subsection{Class-balanced accuracy and response frequencies}
Pairwise balanced accuracy is the mean recall over the three true labels (left, right, tie), which occur 106, 98, and 54 times. A predictor independent of the true label has expected balanced accuracy $1/3$, even if its answer frequencies are unequal. This is the chance reference for balanced accuracy.

\begin{table}[htbp]\centering
\caption{Pairwise performance and predicted-answer frequencies (percent), computed from Table~\ref{tab:tieconfusion}.}\label{tab:balanced}
\begin{tabular}{lrrrrr}\toprule
Checkpoint & Raw accuracy & Balanced accuracy & Left & Right & Tie\\\midrule
Qwen2.5 1.5B & 21.71 & 23.88 & 18.60 & 19.38 & 62.02\\
Qwen2.5 3B & 41.09 & 35.74 & 66.28 & 15.89 & 17.83\\
Qwen3 4B & 44.57 & 36.72 & 94.96 & 3.88 & 1.16\\
Phi-4 mini & 27.52 & 35.73 & 15.89 & 3.49 & 80.62\\
Gemma-3 4B & 57.75 & 48.01 & 79.07 & 19.77 & 1.16\\
Llama-3.2 3B & 24.81 & 32.50 & 17.83 & 7.75 & 74.42\\
\bottomrule\end{tabular}\end{table}
Qwen3's 95.0\% first-alternative rate explains why its non-tie accuracy alone overstates balanced comparison performance. Gemma has the largest balanced-accuracy point estimate, despite zero tie recall. These responses reflect both position preference and performance on count comparisons.

\subsection{Within-checkpoint reading-feature variation}\label{app:featurevariation}
The augmented selector retains all 20 features in the reading block. Eighteen are derived from the model's responses to a request's reading prompts. The other two---the test-coverage flag and development-category reliability---are determined by support counts and checkpoint identity once the development lookup is fixed. Table~\ref{tab:featurevariance} reports all 20, but we exclude the two input-derived covariates when summarizing response variation. We use feature moments from the 20,556 evaluation rows. For each feature separately, let $x_{dum}$ denote its value and assign weight $w_{dum}=1/(4\cdot6\cdot n_d)$. The total weighted variance is $\sum w_{dum}(x_{dum}-\bar x)^2$. The fraction explained by differences between checkpoint means is
\[
\eta^2_{\mathrm{checkpoint}}
=\frac{\frac16\sum_m(\bar x_m-\bar x)^2}
{\sum_{d,u,m}w_{dum}(x_{dum}-\bar x)^2}.
\]
This partitions observed variance by checkpoint. We also center within domains before partitioning checkpoint differences, and report the residual variance within domain/checkpoint cells as a share of total variance. Constant features would be undefined; all 20 have nonzero total variance.

\begin{table}[htbp]\centering\small
\caption{Reading-feature variance decomposition (percent). Checkpoint is the marginal share; conditional is the within-domain checkpoint share; residual is within-domain/checkpoint variance divided by total variance. Cells counts nonconstant domain/checkpoint cells out of 24. Denominators differ across the three percentage columns.}\label{tab:featurevariance}
\begin{tabular}{lrrrr}\toprule
Feature & Checkpoint & Conditional & Residual & Cells\\\midrule
Digits: valid mass & 9.52 & 13.11 & 82.28 & 24\\
Digits: entropy & 28.85 & 46.59 & 51.41 & 24\\
Digits: correct & 4.86 & 6.92 & 89.66 & 22\\
Words: valid mass & 10.43 & 14.03 & 78.60 & 24\\
Words: entropy & 43.61 & 57.15 & 42.70 & 24\\
Words: correct & 5.59 & 7.54 & 86.93 & 23\\
Rows: valid mass & 1.65 & 2.63 & 77.29 & 24\\
Rows: entropy & 51.56 & 56.68 & 42.54 & 24\\
Rows: correct & 0.95 & 1.50 & 85.05 & 24\\
Direct agreement & 18.06 & 25.42 & 73.12 & 24\\
All direct correct & 2.34 & 3.43 & 79.54 & 24\\
Pair: valid mass & 10.56 & 29.21 & 70.49 & 24\\
Pair: correct & 8.58 & 21.21 & 78.50 & 24\\
Reversed pair: valid mass & 4.83 & 19.41 & 80.16 & 24\\
Reversed pair: correct & 3.96 & 16.23 & 83.11 & 24\\
Pair-order agreement & 13.42 & 20.86 & 78.68 & 24\\
Zero/one check: valid mass & 78.36 & 78.39 & 21.61 & 24\\
Zero/one check: correct & 50.15 & 50.18 & 49.82 & 12\\
\midrule
\multicolumn{5}{l}{\textit{Input-derived covariates (excluded from the response-feature summary)}}\\
Development-category reliability & 31.67 & 44.91 & 53.77 & 24\\
Test-coverage flag & 0.00 & 0.00 & 56.94 & 24\\
\bottomrule\end{tabular}\end{table}
For the zero/one check, counts and wording are fixed, and requests differ only in the displayed candidate letters. Its two response features measure stability under letter substitutions in a constant numerical comparison.

Among the 18 response-derived features, the median checkpoint share is 9.98\%; three exceed 50\%, and each varies within at least 12 cells. The excluded covariates supply no new per-request measurements given $V$ and the fixed development lookup, although their explicit encoding can affect a finite-capacity forecaster. Even for the 18 retained measurements, within-checkpoint variation alone does not measure incremental information beyond $V+S_6$.

\section{Checkpoint and family heterogeneity}\label{app:heterogeneity}
Family summaries and leave-family exclusions aggregate the same fitted outcomes, without refitting the selectors. Family intervals fix the domains and resample users within them.

\begin{table}[htbp]\centering
\caption{Family-specific aggregate effects and pooled effects after leaving out the family. Qwen contains three checkpoints; other families contain one each.}\label{tab:family}
\begin{tabular}{lrrr}\toprule
Family & Family effect & 95\% interval & Pooled without family\\\midrule
Qwen & $-0.000198$ & $[-0.003039,0.002573]$ & $-0.003565$\\
Phi & $-0.003007$ & $[-0.007133,0.001168]$ & $-0.001656$\\
Gemma & $-0.006795$ & $[-0.011424,-0.002288]$ & $-0.000899$\\
Llama & $-0.000893$ & $[-0.004848,0.003069]$ & $-0.002079$\\
\bottomrule\end{tabular}\end{table}

\begin{table}[htbp]\centering
\caption{Checkpoint-level accuracy and selection effects. Each effect averages the same four domain means.}\label{tab:checkpoint_effect}
\begin{tabular}{lrr}\toprule
Checkpoint & Test accuracy & Selection effect\\\midrule
Qwen2.5 1.5B & 26.17\% & $0.001303$\\
Qwen2.5 3B & 41.21\% & $0.000316$\\
Qwen3 4B & 58.59\% & $-0.002212$\\
Phi-4 mini & 54.69\% & $-0.003007$\\
Gemma-3 4B & 71.48\% & $-0.006795$\\
Llama-3.2 3B & 48.63\% & $-0.000893$\\
\bottomrule\end{tabular}\end{table}

The Spearman correlation is $-0.942857$. The six checkpoints share users and jointly fitted selectors, while family and size are entangled with test accuracy. Excluding Gemma roughly halves the pooled magnitude. We therefore interpret the correlation as a summary of this model set, without a causal or scaling interpretation.

\FloatBarrier
\Needspace{28\baselineskip}
\section{Analysis coverage and provenance limits}\label{app:coverage}
Table~\ref{tab:coverage} distinguishes analyses in the analysis plan from post hoc additions. The analysis plan is included in the supplement.

\begin{table}[htbp]\centering
\caption{Coverage of planned and post hoc analyses.}\label{tab:coverage}
\begin{tabularx}{\textwidth}{@{}p{.28\textwidth}X@{}}\toprule
Item & Status\\\midrule
Primary, MSE, both rules & Both decision rules fail.\\
Hit@1, all four feature sets & Reported for all four feature sets, including the equal-prompt-count comparison; added interval specifications labeled secondary.\\
Fixed references, oracle gap closed & H/T/F, development-selected fixed policy, interface frequencies and oracle gap closed reported; ratio and best-fixed comparison are post hoc specifications.\\
Evidence specificity & Both random-support ranking and the distinct saved random-evidence prompt control reported.\\
Reading tests and calibration & Overall/difficulty category accuracy, ECE/Brier, actual tie-response cross-tab, forecast calibration and real-request difficulty counts reported.\\
Conventional baselines & Popularity, support, user-KNN and five-seed BPR on identical candidates.\\
Sensitivity analyses & Candidate/development checks reported; fixed-domain interval and the matched-configuration fit labeled post hoc.\\
Cost and reproducibility & Primary, secondary and new comparator intervals have user-array replay; missing forward-pass/timing records limit cost completeness.\\
Construction and chronology & Construction and Last.fm date exposure traced; saved-user timestamp-exclusion sensitivity reported with both pooled interval schemes. The timing and rationale of the KuaiRec threshold change, and whether and when earlier summaries were consulted, remain unknown.\\\bottomrule
\end{tabularx}\end{table}

\FloatBarrier
\section{Statistical replay and computation}\label{app:replay}

\subsection{Statistical replay and refit portability}
Saved-outcome replay reproduces the primary NDCG@5 difference, its 95\% interval and one-sided test, and the domain-level 95\% and 90\% intervals used to evaluate the decision rules. Linux execution records also report byte-identical evaluation and additional-candidate rows after feature construction and forecaster refitting: Python 3.12.13, NumPy 2.4.4, scikit-learn 1.8.0, and SciPy 1.17.1, using two CPU threads with no GPU visibility. Three development-MSE entries differ by at most $2.78\times10^{-17}$, while output rows and selected configurations agree.

A separate macOS arm64 refit yields $-0.001972$ with interval $[-0.004453,0.000252]$ and one-sided $p=0.968303$, so refitting is not bitwise portable. Direct statistical replay remains exact. The exact difficulty-mapping code and amendment are included. Appendices~\ref{app:promptvariants} and~\ref{app:zeroone} specify the prompts; \texttt{PROMPT\_SPECIFICATION.md} supplies the defining source references. The replay starts from saved predictions; regenerating LLM outputs requires the upstream data and inference pipeline.

\subsection{Prompt requirements and computation}
The plan defines the primary comparison as adding reading features to the baseline selector, although its cost paragraph describes equal diagnostic budgets. The supplement retains the plan and its reporting correction. Table~\ref{tab:budget} gives the corrected requirements.

\begin{table}[htbp]\centering
\caption{Required prompts per checkpoint--request; these are feature requirements rather than separate timed policy executions. The experiment scores the full prompt collection before policy selection.}\label{tab:budget}
\begin{tabular}{lrr}\toprule
Boundary & Baseline & Baseline + reading\\\midrule
Shared main H/T prompts & 8 & 8\\
Additional diagnostic prompts & 6 & 12\\
Before shared random-control scoring & 14 & 20\\
Including four shared random-control prompts & 18 & 24\\\bottomrule
\end{tabular}\end{table}

The development/evaluation recommendation ledger records 759,456 prompts and 321,023,045 unpadded input tokens across six checkpoints. It excludes other stages, including complete reading-test and additional-candidate runs. With one partial-session record missing, retained sessions give lower bounds of 79,089 forward passes, 476,085,792 padded input tokens, and about 50 hours of shared-GPU forward time; the largest allocated-memory peak is 20.6 GiB. These measurements are neither complete run totals nor isolated per-policy latency. Scoring uses next-token logits without generated completions.

Statistical replay also succeeds under Python 3.13.5 and NumPy 2.3.5. Receipts identify each environment; this portability concerns saved-outcome statistics, not model refitting.

\subsection{Statistical artifact boundary}
One scalar per user reproduces the primary; further checkpoint-averaged arrays reproduce feature-set comparisons, Hit@1, MSE, fixed-policy references, and choice frequencies. The control/matched-fit array supplies 21 paired metrics for all 3,426 users. Its fixed-domain replay exactly matches 234 numeric targets, including interval endpoints and resampling settings, in the supplied Linux and local receipts. The matched-fit outcomes were recovered by replaying the already specified CPU fit, without further configuration search.

Per-cell feature moments and confusion counts reproduce the diagnostic summaries; the reporting checker separates response features from input-derived covariates. The source-date audit supplies aggregate results and an upstream-dependent scan script. Its aligned 426-entry mask and user-ID-join receipt support the restriction in Appendix~\ref{app:timestamp_sensitivity}; statistical replay checks the array hashes and group accounting without repeating the source joins. Histories, user-specific prompts, source identifiers, and full model outputs are excluded.

\end{document}